\documentclass[pre,twocolumn,amsmath]{revtex4-1}
\pdfoutput=1
\usepackage{graphicx}

\usepackage{amssymb,amsfonts,amsmath}

\def \ra{\rightarrow}
\def \>{\rangle}
\def \<{\langle}

\def\be{\begin{equation}}
\def\ee{\end{equation}}
\def\longrightharpoonup{\relbar\joinrel\rightharpoonup}
\def\longleftharpoondown{\leftharpoondown\joinrel\relbar}

\def\longrightleftharpoons{
  \mathop{
    \vcenter{
      \hbox{
      \ooalign{
        \raise1pt\hbox{$\longrightharpoonup\joinrel$}\crcr
	  \lower1pt\hbox{$\longleftharpoondown\joinrel$}
	  }
      }
    }
  }
}

\newcommand \bea {\begin{eqnarray}}
\newcommand \eea {\end{eqnarray}}

\begin{document}

\title{ Environmental engineering is an emergent feature of diverse ecosystems and drives community structure}

\author{Madhu Advani}
\affiliation{Harvard University,Cambridge, MA 02142}

\author{Guy Bunin}
\affiliation{Technion-Israel Institute of Technology, Haifa, 3200003, Israel}

\author{Pankaj Mehta}
\email{pankajm@bu.edu}
\affiliation{Dept. of Physics, Boston University, Boston, MA 02215}

\begin{abstract}

A central question in ecology is to understand the ecological processes that shape community structure. Niche-based theories have emphasized the important role played by competition for maintaining species diversity. Many of these insights
have been derived using MacArthur's consumer resource model (MCRM) or its generalizations. Most theoretical work on the MCRM has focused on small ecosystems with a few species and resources. However theoretical insights derived from small ecosystems many not scale up large ecosystems with many resources and species because large systems with many interacting components often display new emergent behaviors that cannot be understood or deduced from analyzing smaller systems. To address this shortcoming, we develop a sophisticated statistical physics inspired cavity method to analyze MCRM when both the number of species and the number of resources is large.  We find that in this limit, species generically and consistently perturb their environments and significantly modify available ecological niches. We show how our cavity approach naturally generalizes niche theory to large ecosystems by accounting for the effect of this emergent environmental engineering on species invasion and ecological stability. Our work suggests that environmental engineering is a generic feature of large, natural ecosystems and must be taken into account when analyzing and interpreting community structure. It also highlights the important role that statistical-physics inspired approaches can play in furthering our understanding of ecology.

 \end{abstract}

\maketitle

\section{Introduction}
One of the most stunning aspects of the natural world is the diversity of species present in most ecosystems. The community structure of ecosystems are shaped through a complex interplay of the externally supplied resources available in an ecosystem, competition for these resources, as well as stochasticity  \cite{chesson2000mechanisms, tilman1982resource, vellend_conceptual_2010, hubbell_unified_2001}. A fundamental problem in community ecology is to understand how these processes give rise to observed pattern of species abundances. A rich theoretical framework has been developed to address this problem. Niche-based theories have emphasized the role of competition for resources \cite{hardin_competitive_1960, macarthur1967limiting,macarthur1970species, chesson_macarthurs_1990, tilman1982resource, chase_ecological_2003, letten2017linking}, while neutral theory has highlighted the role of stochastic effects \cite{hubbell_unified_2001, volkov_neutral_2003, rosindell_unified_2011, rosindell_case_2012}, and several works have investigated the interplay between stochasticity and competition \cite{chisholm2011theory, fisher2014transition, jeraldo2012quantification, tilman_niche_2004, gravel_reconciling_2006}.

Many of these theoretical insights have been synthesized in what is commonly referred to as contemporary niche theory. Contemporary niche theory highlights the role played by equalizing mechanisms, processes that decrease fitness differences between organisms, and stabilizing mechanisms, processes that decrease competition for resources. These basic organizational schema have been successfully applied to understand community structure in a wide range of settings \cite{chesson2000mechanisms, tilman1982resource, vellend_conceptual_2010}.

One of the simplest and most influential mathematical models for niche theory is MacArthur's consumer resource model (MCRM) \cite{macarthur1970species, chesson_macarthurs_1990, tilman1982resource, letten2017linking}. Most analysis of  MCRM -- including those that inform contemporary niche theory and modern coexistence theory -- have focused on small ecosystems with a few species and and few resources \cite{macarthur1970species, chesson_macarthurs_1990, tilman1982resource, letten2017linking}. However, it is unclear to what extent the theoretical insights derived from ecosystems with just a few species can be scaled up to  diverse, natural ecosystems. One of the defining features of large complex systems is that they often display new ``emergent behaviors'' that cannot be understood or deduced from analyzing small systems with just a few parts \cite{anderson1972more, levin1992problem, levins1966strategy, levins1985dialectical}. For this reason, it is essential to directly analyze large ecosystems with many resources and species and ask how they differ from the few-species ecosystems that have been analyzed previously.  Recently, several works suggest that large ecosystems can exhibit unexpected behaviors such as phase transitions, emergent community-level cohesion, and the analogues of critical points \cite{fisher2014transition, dickens2016analytically, tikhonov2017collective, kessler2015generalized, tikhonov2016community, bunin2016interaction}. This highlights the need for new theoretical frameworks for directly analyzing large, heterogeneous ecosystems.

Perhaps the most successful and ubiquitous approaches for analyzing large systems in statistical physics is mean field theory. We emphasize that what is meant by a mean field theories in statistical physics is distinct from the way it is commonly understood in ecology \cite{kadanoff_statistical_2000, violle2012return}. Unlike most usages in ecology, mean field theories in physics account for not only the means of various quantities but also fluctuations around the mean. In this paper, whenever we use the term mean field theory, we will mean it in this broader statistical physics definition rather than the narrow usage common in ecology. Mean field models have long history in statistical physics and have played a central role in the study of phase transitions and collective emergent behaviors in physical systems \cite{kadanoff2000statistical, landau1980statistical}. Most mean field theories in physics focus on homogenous systems with identical components and couplings. However, more sophisticated variants such as the cavity method can be used to analyze heterogeneous ``disordered systems'' \cite{opper2001advanced}. Here, we develop a statistical physics inspired mean field theory, based on a generalization of the cavity method, and use it to analyze diverse ecosystems. In this paper, we will refer to this as the cavity theory (CT).

Our methods are inspired by and build upon recent work showing the connection between community ecology the physics of disordered systems \cite{diederich1989replicators, eissfeller1992new,rieger1989solvable,tokita2004species, fisher2014transition, dickens2016analytically, tikhonov2017collective, bunin2016interaction,barbier2017generic, yoshino2007statistical}. It is also closely related to the statistical mechanics of interacting socio-economic agents \cite{de2006statistical}. However, unlike these previous works our analysis explicitly incorporates  resource dynamics, including resource heterogeneity and depletion. This allows us to naturally connect our results to contemporary niche theory and modern coexistence theory.  One of the most striking aspects of our analysis is that we find that environmental engineering is a generic feature of all diverse ecosystems \cite{erwin2008macroevolution}. In diverse ecosystems, organisms can and do significantly reshape their environments by changing resource abundances and, importantly, depleting resources. Moreover, we show that many of the central theoretical quantities in our novel CT have natural ecological interpretations that generalize many classical quantities and results of niche theory to large ecosystems and quantify the role of environmental engineering in shaping community structure.

\begin{figure}[t]
\includegraphics[width=1.0\columnwidth]{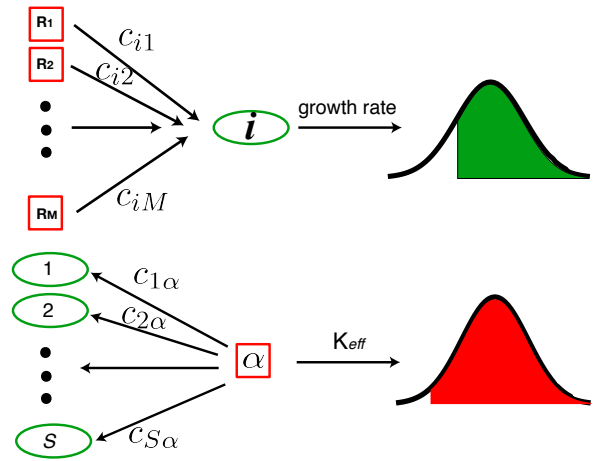}
\caption{{\bf Analyzing MacArthur Consumer's Resource Model for large, diverse ecosystems.}  (Top) The growth rate
for a species $i$ is a sum of  terms resulting from consuming  $M \gg 1$ resources. For this reason, from the central limit theorem, it can be well modeled by a (truncated) normally distributed variable. (Bottom) Each resource $\alpha$ is consumed by $S \gg 1$ consumers. From the central limit theorem, the effective carrying capacity of the resource (i.e. the
resource abundance at steady-state) can also be modeled using a  (truncated) normal distribution. The truncation is due to the fact that neither species nor resource abundances can become negative. To derive our analytic cavity equations, we require self-consistency for the means and variance of these distributions. }
\end{figure}

\section{MacArthur consumer resource model}

In this work, we will analyze one of the canonical and most influential models in community ecology: MacArthur's Consumer Resource Model (MCRM) \cite{macarthur1970species, chesson_macarthurs_1990}.  MCRM consists of $S$ species or consumers with abundances $N_i$ ($i=1 \ldots S$)  that can consume one of $M$ substitutable resources with abundances $R_\alpha$ ($\alpha=1 \ldots M$). The consumer preferences of species $i$ for resource $\alpha$ are encoded by a $S \times M$ matrix, $c_{i \alpha}$.

In the MCRM, the growth rate $g_i(\mathbf{R})$ of a species depends of the concentration of all the resources. To model the growth rate, following MacArthur, we assume that a species $i$ have some minimum maintenance cost, $m_i$, that they must meet. The growth rate, $g_i(\mathbf{R})$,  is proportional to amount of resources consumed, weighted by a quality factor $w_\alpha$, minus this maintenance cost
\be
g_i(\mathbf{R})= \sum_\alpha c_{i \alpha} w_\alpha  R_\alpha -m_i.
\label{Eq:growth_rate}
\ee
If $g_i>0$, then this is also the growth rate of species $i$.

The resources have their own internal dynamics which, following MacArthur, we assume can be modeled using logistic growth. Furthermore, when a resource is consumed, it's abundance is reduced. This ecological dynamics is captured by the following coupled, nonlinear differential equations
\bea
{dN_i \over dt} &=& N_i g_i(\mathbf{R}) \nonumber \\
{dR_{\alpha} \over dt}&=& F_\alpha (R_\alpha) - \sum_i N_i c_{i \alpha} R_\alpha,
\label{Eq:MCRM-original}
\eea
where $F_\alpha(R_\alpha)=R_\alpha(K_\alpha -R_\alpha) $ describes the resource dynamics in the absences of consumption and $K_\alpha$ is the carrying capacity of each resource $\alpha$. In our model, both the species and resource abundances $N_i$ and $R_\alpha$ must be strictly positive. For our analysis, it will be useful to define an ``effective resource capacity''
\be
K_\alpha^{eff} (\mathbf{N}) =K_\alpha - \sum_i N_i c_{i \alpha}
\label{Eq:Keff}
\ee
 that accounts for depletion of resources by consumers \cite{dickens2016analytically}. The MCRM can be rewritten in terms of  $K_\alpha^{eff} (\mathbf{N})$ as
 \bea
{dN_i \over dt} &=& N_i g_i(\mathbf{R}) \nonumber \\
{dR_{\alpha} \over dt}&=& R_\alpha(K_{\alpha}^{eff}(\mathbf{N}) -R_\alpha)
\label{Eq:MCRM}
\eea
A crucial property of these equations is that resources can be completely depleted from the environment. This will play an important role in what follows. Finally, we emphasize that these equations are identical those analyzed by MacArthur, Chesson, and others in deriving modern niche theory.

\section{ Statistical mechanics approach to MacArthur's Consumer Resource Model}

\begin{figure}[t!]
 \includegraphics[width=1.0\columnwidth]{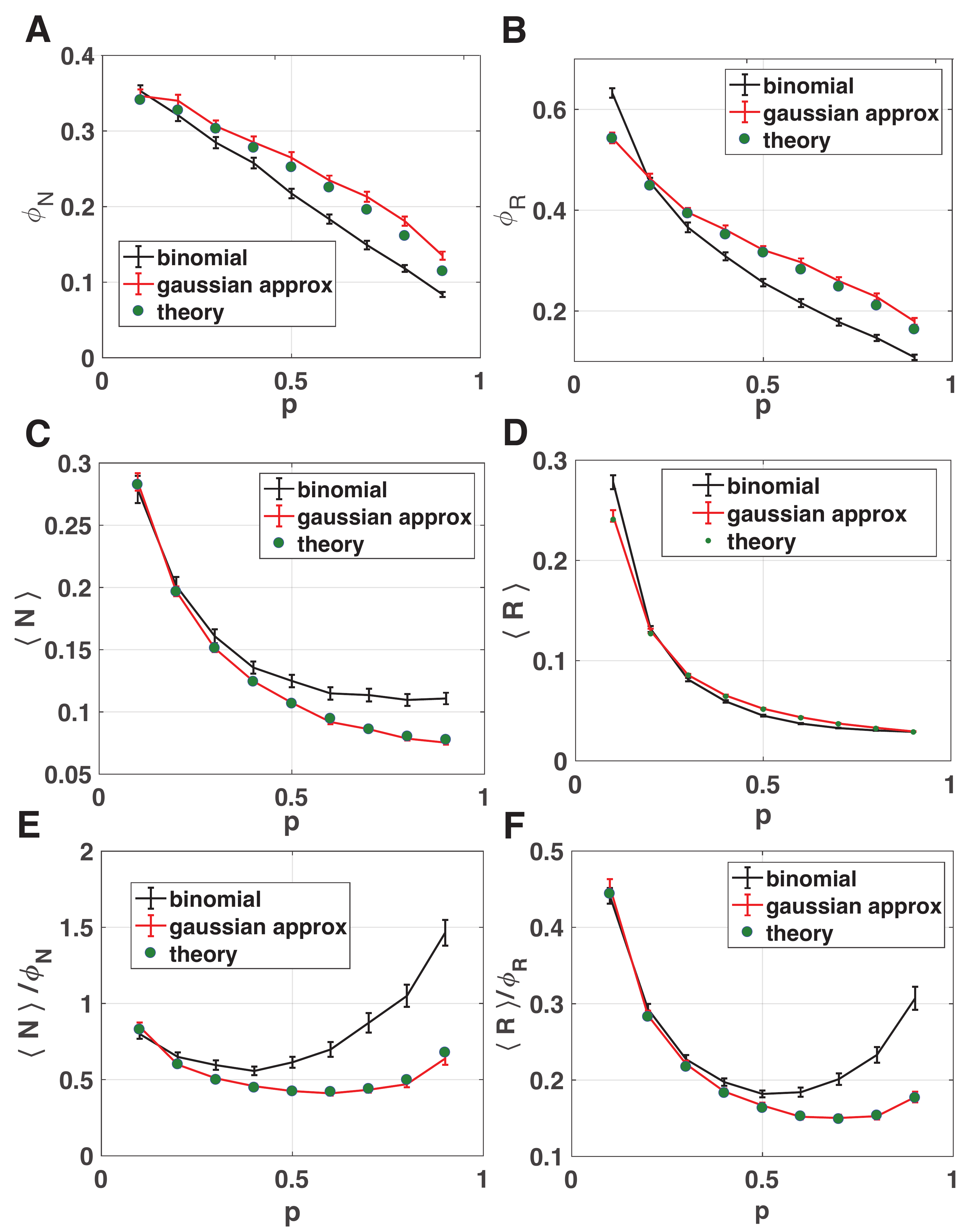}
\caption{{\bf Comparison of numerical simulations with theory}. Ecosystems were simulated with the consumption coefficients $c_{i \alpha}=0,1$ drawn from a Bernoulli distribution
with probability $p$ (black lines) or a Gaussian distribution with same mean and variance as the binomial distribution (red line). Here $S = M = 30$, $K=1$, $\sigma_K=1$, $m=1$, $\sigma_m=.1$,  and 250 trials were used in these simulations: the error bars denote $\pm 2 $ standard deviations. The $K_a$ and $m_i$ were drawn iid from a gamma distribution in the binomial plot to ensure non-negativity of the parameters, and a Gaussian distribution in the Gaussian approximation plot. The numerical results were compared to theoretical predictions from the
self-consistent cavity equations for ecologically stable steady-states (green circles).  (A) Fraction of species that survive $\phi_N$, (B) fraction of resources that are not depleted $\phi_R$, (C) average abundance of all species $\< N\>$,
(D) average abundance of all resources $\<R\>$ as a function of the probability $p$. (E) Mean-abundance of surviving species  $\< N\>/\phi_N$  and (F) mean-abundance
of surviving resources $\<R\>/\phi_R$.}
\label{CT-numeric1}
\end{figure}

Previous approaches to analyzing the MCRM have largely been confined to small ecosystems with a few species and resources. Here, we consider the opposite limit of large, diverse ecosystems where both the number of species and number of resources is large, $S, M \gg 1$. In this limit, the number of parameters needed to define the ecosystem dynamics becomes extremely large. To overcome this problem, we follow a long tradition in theoretical ecology pioneered by Robert  May of looking at the case where the parameters are drawn from a random distribution \cite{may_will_1972}. This allows us to ask questions about the behavior of a generic, diverse ecosystem.

We consider the case where all the consumption coefficients $c_{i\alpha}$, resource carrying capacities $K_\alpha$,  and maintenance costs $m_i$ are drawn from a random distribution. Our analytic calculations depend only on the mean and variances of the probability distributions. Denoting the expectation value of a parameters $x$ over a distribution by $\<x\>$, then we denote the mean and variances of our parameters by: $\< c_{i \alpha}\> = \mu_c /S$ , $\<  (c_{i \alpha} -\<c_{i \alpha}\>)^2\> = \sigma_c^2 /S$, $\<m_i\>=m$, $\<( m_i -\<m\>)^2\>=\sigma_m^2$, $\<K_\alpha \>=K$, and $\<(K_\alpha-\<K_\alpha\>)^2\>=\sigma_K^2$. We can also define a parameter $\gamma=M/S$ that measures the ratio of resources to species.

\subsection{Invasion, ecological stability, and self-consistency}

One of the cornerstones of community ecology is the idea of invasion \cite{macarthur1967limiting,shea2002community, tilman_resource_1982}.  In our analysis, we will ask under what circumstances a new species can invade an ecosystem. Denote the growth rate of species $i$ when it tries to invade the ecosystem $g_i^{inv}$. We will call this the invasion growth rate.  Since we are interested in statistical properties, we will be primarily concerned with the mean and variances of the invasion growth rate averaged
over all species $i$ in the regional species pool:  $\< g_i^{inv}\>=g$ and $\<(g_i^{inv})^2\>-\<g_i^{inv}\>^2=\sigma_g^2$.

The key idea that we will exploit in our analysis is the observation that as $S$ and $M$ get large, both the invasion growth rates,  $g_i^{inv}$, and the effective carrying capacities, $K_{\alpha}^{eff}$ are the sum of a large number of small terms. Each individual resource makes only a small contribution (of order $1/M$) to the growth of any consumer, and every consumer makes an order $1/S$ contribution to the effective resource capacity. Thus, from the central limit theorem, the distribution of growth rates $g_i^{inv}$ and the distribution of effective resources $K_{\alpha}^{eff}$ in the ecosystem can be well-approximated by a normal distribution. In the language of the cavity method of statistical physics, this corresponds to the replica symmetric solution. For future reference, denote the means and  variance of the effective carrying capacity by $\<K_{\alpha}^{eff}\>=K^{eff}$ and  $\<(K_{\alpha}^{eff})^2\>-\<K_{\alpha}^{eff}\>^2= \sigma_{K^{eff}}^2$ (see Figure 1).

This suggests the following intuition for thinking about our ecosystem. Each species, $i$, has a invasion growth rate drawn from a normal distribution. In other words, we can think of $g_i^{inv} \approx g + \sigma_g z_i$, where $z_i$ is a standard normal variable. Similarly, each resource has an effective carrying capacity that is also drawn from a normal distribution, with $K_\alpha^{eff}\approx K^{eff} + \sigma_K \tilde{z}_\alpha$, with $\tilde{z}_\alpha$ a standard, normal variable. In general, the means and variances ($g,  K^{eff}, \sigma_g^2, \sigma_{K^{eff}}^2$) depend on the abundances of all other species and resources.  Our statistical mechanics inspired mean field approach exploits this observation to self-consistently solve for the means and variances of the invasion growth rate and effective resource carrying capacity. In the physics literature, these is known as cavity theory (CT).  In general, this is a very subtle calculation but can be done using a generalized cavity equation (see below and in appendix).

In order to derive the CT self-consistency equations, we consider a system with $S$ species and $M$ resources and ask what happens when we add an additional species and resource to the system.  We denote the abundances of the additional species and resource by $N_0$ and $R_0$ respectively. This two-step cavity where both a resource and species is removed is similar to the procedure employed to analyze the Hopfield model and compressed sensing \cite{shamir2000thouless, ramezanali2015critical} and is necessary to correctly capture subtle correlations between resource and species dynamics due to environmental engineering. This approach is intimately related to classic works by MacArthur and Levins that analyzed ecological dynamics by asking if a new species could invade an ecosystem \cite{macarthur1967limiting}. Whereas their analysis was applicable to small ecosystems with a few species, our analysis is valid for large, diverse ecosystems.

Since the number of species and resources in the original ecosystem is large ($S,M \gg 1$), the addition of the new resource and consumer represent a small perturbation of the original system. For this reason, it is useful to define two susceptibilities, $\chi$ and $\nu$ that measure the sensitivity of an ecosystem to small perturbations. The resource susceptibility, $\chi$,  measures the average change in the mean resource abundance at steady-state if we slightly increase the supply of all the externally supplied resources. Denoting the steady-state value of a quantity $X$ by $\bar{X}$, we can mathematically define $\chi$, as
\be
\chi = {1 \over M} \sum_\alpha {\partial \bar{R}_\alpha \over \partial K_\alpha}.
\ee
The average species-cost susceptibility, $\nu$, measures the change in mean species abundances if we slightly decrease the minimum fitness cost (or equivalently increase the growth rate),
\be
\nu={1 \over S} \sum_i {\partial \bar{N}_i \over \partial g_i} = -{1 \over S} \sum_i {\partial \bar{N}_i \over \partial m_i}.
\ee
These susceptibilities characterize the sensitivity of an ecosystem to perturbations and can be directly measured in experiments.

In terms of these quantities, one can derive a simple expression for the steady-state abundances of newly added consumer and resource (see Appendix):
\bea
\bar{N}_0  &=& { \max \left[ 0,  {g + \sigma_g {z}_0 }\right] \over  \gamma \sigma_c^2 \chi}, \nonumber \\
\bar{R}_0 &=& { \max \left[0, K^{eff}+ \sigma_{K^{eff}}^2 \tilde{z}_0\right ] \over 1-\sigma_c^2 \nu},
\label{EqNRmain}
\eea
where as above $z_0$ and $\tilde{z}_0$ are independent, unit normal variables. These equations have a beautiful and straightforward interpretation. A new species added to the system will have an invasion growth rate $g_0^{inv}=g + \sigma_g {z}_0 $, which is normally distributed. If the growth rate is negative, it will not be able to invade the system and go extinct. If its growth rate is positive when introduced in the ecosystem, then it survives with an abundance proportional to its invasion growth rate. We emphasize that this proportionality constant can differ significantly from what would be expected in a single-species ecosystem and depends on all the other resources and species present in the ecosystem through the susceptibility $\chi$ and the variance of the consumption coefficients $\sigma_c^2$. For this reason, the invasion growth rate of a species when it invades an ecosystem is positively correlated
with its abundance.  Similarly, the new resource is depleted if its effective carrying capacity is negative. Otherwise  the steady-state abundance of the new resource is proportional to its effective carrying capacity. These equations are similar to the arguments of MacArthur and Levins on the necessary conditions for invasibility to large ecosystems  \cite{macarthur1967limiting}. They also generalize results for species abundances derived in  \cite{bunin2016interaction} using the Lotka-Volterra equation and the results in \cite{ de2006statistical, yoshino2007statistical, tikhonov2017collective} which ignored resource depletion and resource fluctuations.

\subsection{Comparison with numerics}

Unlike small ecosystems,  we cannot analytically solve for the all the resource and species abundances. However, we can take a statistical approach that allows us to calculate statistical properties of species and resource abundances at steady-state. We also restrict our analysis to uninvadable steady-states, defined as a steady-state which cannot be invaded by any species. This, both simplifies the mathematics, and allows us to more directly relate our calculations to ecology.

Using (\ref{EqNRmain}) is it possible to derive self-consistency equations for the fraction of species in the regional species pool that survive, $\phi_N$, the mean abundance of the species $\<N\>=1/S \sum_i N_i$,  and variance and second moment of surviving species abundances, $\<(\delta N)^2\>$ and $q_N= \<(\delta N)^2\>+ \<N\>^2= 1/S \sum_i N_i^2 $ respectively.  We can also calculate the analogous equations for resources: the fraction of resources with non-zero abundance, $\phi_R$, the mean abundance of  resources $\<R\> = 1/M \sum_\alpha R_\alpha$,  and variance and second moment of the resource abundances, $\<(\delta R)^2\>$ and $q_R= \<(\delta R)^2\>+ \<R\>^2 = 1/M \sum_{\alpha} R_\alpha^2$. The equations are derived in Appendix \ref{sec:self-con} and can be solved numerically.

To check the accuracy of our CT, we compared our analytic predictions to numerical simulations (see Figure \ref{CT-numeric1}). We simulated  (\ref{Eq:MCRM-original}) for two different choices of distributions for the $c_{i \alpha}$.  In the first set of simulation,  the $c_{i \alpha}$ were binary random variables with  $c_{i \alpha}=1$ with probability $p$ and $c_{i \alpha}=0$ with probability $1-p$. The probability $p$ can be viewed as the level of generalism in the regional species pool. As $p \rightarrow 0$, all organisms in the community are specialist and consume a handful of resources. When $p \rightarrow1$, the community consists of generalists who can consume almost all resources. In the second set of simulations, we drew the consumption coefficients from a Gaussian distribution with the same mean and variance as the corresponding Bernoulli distribution with probability $p$.

As shown in Fig. \ref{CT-numeric1}, our analytic results agree remarkably well with numerical simulations. The agreement between theory and numerics is nearly exact when $c_{i \alpha}$ are drawn from a Gaussian and shows qualitative agreement even when the consumption coefficients $c_{i \alpha}$ are binary random variables. This is a result of the Gaussianity assumptions used to derive the cavity equations (see Appendix). The discrepancy between the binary case and Gaussian case stems from the fact that the for large $S$ and $M$ the $c_{i \alpha}$ are strictly positive for the binary case but generically contain some negative elements for Gaussian distributions. A negative $c_{i \alpha}$ implies that species $i$ produces resource $\alpha$ at a fitness cost to itself. Thus, all simulations with Gaussian include a small fraction of public good producers that are accounted for in our theoretical calculations but are absent in the simulations with binary variables.

Despite these differences, for both choice of distributions the fraction of surviving species declines with increasing $p$. This is consistent with the basic idea of niche-theory that as $p$ increases, there is increased competition resulting in greater competitive exclusion. In contrast,  the mean abundances of surviving species and resources shows a non-monotonic behavior as a function of $p$ in both numerical simulations and analytics (see appendix and Fig. \ref{fig:app_gauss_crm} for additional simulation results).

 \begin{figure}[t!]
 \includegraphics[width=1.0\columnwidth]{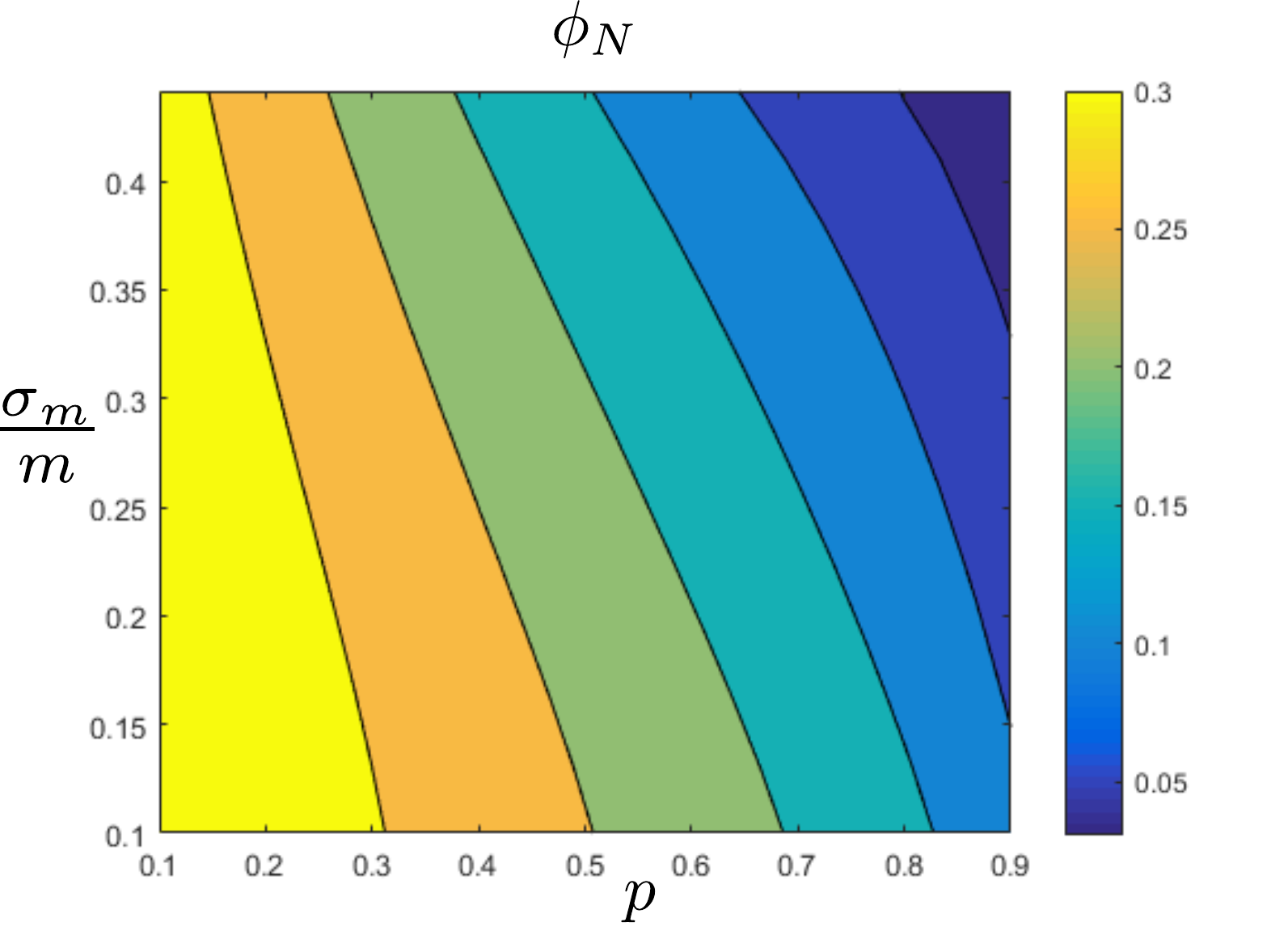}
\caption{{\bf Co-existence and diversity.} Here we use the same parameters as in the previous plot, apart from allowing $\sigma_m$ to vary, and show the cavity prediction of the fraction of surviving species $\phi_N$ as a function of $p$ and the standard deviation over mean $\sigma_m/m$ of the maintenance costs $m_i$ of species. In this regime increasing $p$ leads to more similar species by increasing the niche overlap $\rho$ as defined in \eqref{Eq:nicheoverlap} from $\rho \approx .77$ to $\rho \approx .99$ in the range shown above. As predicted by niche theories, increasing $\rho$ leads to increased competition and a smaller $\phi_N$ while decreasing $\sigma_m/m_i$ leads to larger fraction of species surviving at a fixed $p$. }
\label{niche-phasediagram}
\end{figure}

%\caption{{\bf Co-existence and niche overlap.} The fraction of surviving species $\phi_N$ as a function of $p$ and the standard deviation over mean $\sigma_m/m$ of the maintenance costs $m_i$ of species.
%As predicted by niche theories, increasing $\rho$ leads to increased competition and a smaller $\phi_N$ while decreasing $\sigma_m/m_i$ leads to larger fraction of species surviving at a fixed $p$. }

\section{Generalizing niche theory to large ecosystems}

The MacArthur consumer resource model has played a central role in the development of niche-based theories of community assembly \cite{macarthur1970species, chesson_macarthurs_1990, tilman1982resource, chase_ecological_2003, letten2017linking}. However, most of these analyses have focused on small ecosystems with just a few species and resources. Here, we discuss the ecological implications of our analysis for understanding community assembly in large ecosystems with many species and resources.

\subsection{Relating MCRM parameters to ecology}

We begin by relating the parameters of the MCRM to more ecologically meaningful quantities such as the niche overlap, fitness, zero net- growth isoclines (ZNGI), and impact vectors. In ecology, the niche overlap, $\rho$, measures how much two species compete for the same resources. The larger the niche overlap, the more species compete. For small ecosystems, the niche overlap is  bounded between $0$ and $1$, with a niche overlap of zero meaning the species do not compete for resources and a niche overlap of one indicating the species have identical consumption profiles. In the context of the two species MacArthur resource model, the niche-overlap between species can be thought of as the percentage of variance explained if one performs a regression of the first consumer's consumption vector against the consumption vector  of the second species\cite{macarthur1970species, chesson_macarthurs_1990, chesson2000mechanisms}.  Using this observation, we can naturally extend the idea of niche overlap to entire ecosystems by defining an ecosystem-level niche overlap $\rho$ in terms of the mean and variances of the consumption coefficients $c_{i \alpha}$:
\be
\rho = { \mu_c ^2  \over {\mu_c^2 +\sigma_c^2}}.
\label{Eq:nicheoverlap}
\ee
One useful way of thinking about $\rho$ is that it measures the niche-overlap between two species randomly drawn from the regional species pool.  It is easy to see that when $\sigma_c^2 \ll \mu_c$, all species have nearly identical consumption preferences and $\rho \rightarrow 1$. In contrast when $\sigma_c^2 \gg \mu_c$, species will have very distinct consumer preferences and $\rho \rightarrow 0$.

Another fundamental quantity in contemporary niche theory is the ecological fitness of an organism, $f_i= \sum_{\alpha} c_{i \alpha} K_\alpha -m_i$ \cite{chesson_macarthurs_1990, chesson2000mechanisms}. This fitness is the initial growth rate of organism $i$ in the  \emph{absence} of other species. In general, the actual growth rate of a species will differ significantly from the fitness if the resource abundances differ significantly from the resource carrying capacities $K_\alpha$. For this reason, we will refer to this as the ``naive'' fitness.

We show in the appendix that it is also possible to relate our parameters directly to ZGNIs  and generalized impact vectors.

\subsection{ Niche overlap and coexistence}

One of the fundamental results of niche-based theories is that as the niche-overlap between species increases, co-existence become more and more difficult \cite{chesson2000mechanisms}. The underlying reason for this is species that have similar consumer preference are more likely to compete with each other, resulting in competitive exclusion. Thus, increasing the niche-overlap in the community should decrease the fraction of species  $\phi_N$ that can co-exist in a community. On the other hand, stabilizing mechanisms that decrease the fitness differences between species should increase coexistence.  We can parameterize the fitness differences in the community by the dimensionless quantity $\sigma_m/m$  equal to the standard deviation over the mean of the maintenance costs $m_i$ over all species in the regional species pool. This choice of parameterization is in line with contemporary niche theory where fitness differences are defined as the difference in growth rates when species have identical consumption preferences  \cite{chesson2000mechanisms}. Figure \ref{niche-phasediagram} shows $\phi_N$ as a function of the niche overlap $\rho$ and $\sigma_m^2/m$. This choice of niche-overlap corresponds to varying the probability $p$ for having a non-zero $c_{i \alpha}$ from $0.1$ to $0.9$  (see Fig. \ref{CT-numeric1}). As predicted by niche theories, increasing $\rho$ leads to increased competition and a smaller $\phi_N$. In constrast decreasing $\sigma_m/m_i$ at a fixed $\rho$, leads to a larger fraction of species surviving. Thus, in this regard large ecosystems behave quite similarly to predictions made by analyzing smaller models.

\begin{figure}
\includegraphics[width=1.0\linewidth]{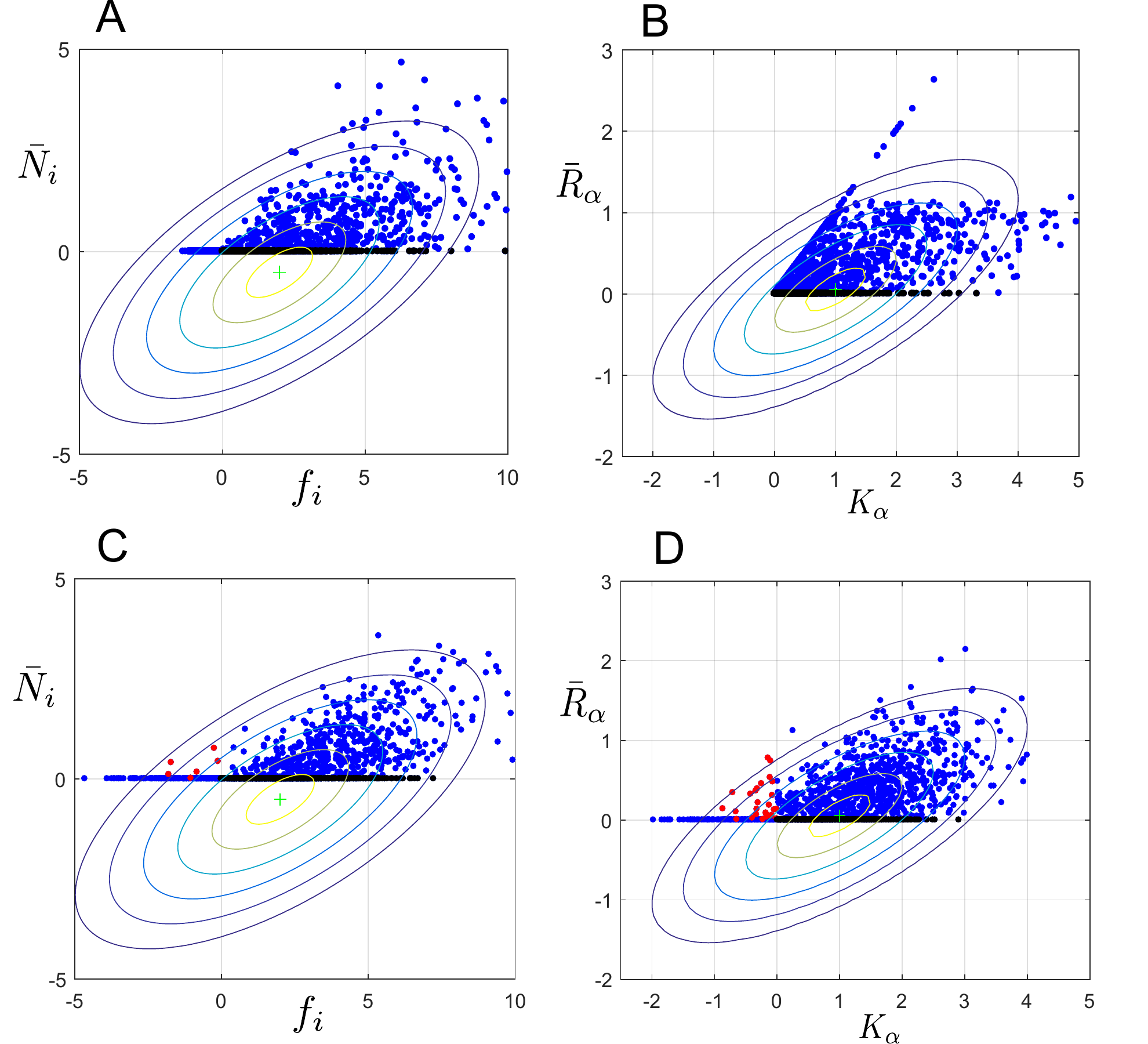}
\caption{{\bf Measuring the effect of environmental engineering} (A) Steady-state abundance $\bar{N}_i$ versus the fitness  $f_i=\sum_{\alpha}c_{i \alpha}\max\{K_\alpha,0\} -m_i$ for each species $i$ . Fitness is defined as the initial growth rate of species $i$ in the environment in the absence of all other species. Points colored black are species with positive fitness that go extinct in the community ($f_i>0$, $\bar{N}_i=0$). (B) Comparison of the steady state resource levels $\bar{R}_\alpha$ with their capacity $K_\alpha$. The filled circles are generated from simulations with M = S = 30 resources and species, the data is generated from 50 separate trials. Parameters for simulations as in Figure \ref{CT-numeric1} with $p=0.1$ and $\sigma_m/m=0.1$.. Black points indicate resources which have a positive capacity but go extinct in the community. The difference between plots for (A,B) and (C,D) is that in the former $K_\alpha$ and $m_i$ are always positive because they are drawn from a gamma distribution and Bernoulli distribution respectively. In (C,D), each of these parameters is drawn from a Gaussian distribution with the same mean and variance as in (A,B). This means that a small fraction of the $K_\alpha, m_i$, and $c_{i\alpha}$ are negative. Negative $c_{i \alpha}$ corresponds to production of resource $\alpha$ by species $i$ at a fitness cost to itself (i.e. public good production). In (C,D), red points indicate species with negative fitness that can stably exist in the community  ($f_i <0$, $\bar{N}_i>0$) or resources with negative $K_\alpha$ that are produced by the ecosystem. Contours show theoretical predictions of our CT for correlation between $\bar{N}_i$ and $f_i$ as well as $\bar{R}_\alpha$ and $K_\alpha$ (see Appendix \ref{sec:f_N_thy} for details). Each contour represents half a standard deviation of our theory.}
\label{fig:Nversusf}
\end{figure}

%{\bf Measuring the effect of environmental engineering}. A) Steady-state abundance $\bar{N}_i$ versus the fitness $f_i=\sum_{\alpha}c_{i \alpha}K_\alpha -m_i$ for each species $i$.  Fitness is defined
%as the initial growth rate of species $i$ in the environment in the absence of all other species. Points colored black are species with positive fitness that go extinct in the community ($f_i>0$, $\bar{N}_i=0$). Points colored red indicate species with negative fitness  that can stably exist in the community ($f_i <0$, $\bar{N}_i>0$).  B) Comparing the steady state resource levels $\bar{R}_\alpha$ with their capacity $K_\alpha$ and matching these with the cavity predictions. In this case, points which are colored black are resources which have a positive resource capacity but end up extinct in the environment, and those colored red have a negative resource capacity but end up surviving in the environment. Both plots demonstrate that environmental engineering can dramatically modify community structure. Contours show theoretical predictions of our CT for correlation between $\bar{N}_i$ and $f_i$ as well as $\bar{R}_\alpha$ and $K_\alpha$ (see Appendix \ref{sec:f_N_thy} for details). In both cases, the filled circles are generated from simulations with $M = S= 30$ resources and species, the data is generated from $50$ separate trials. In both plots, each contour represents half a standard deviation of our Gaussian fit. Parameters for simulations as in Figure \ref{CT-numeric1} with $p=0.1$ and $\sigma_m/m=0.1$. }

\subsection{Resource depletion and environmental engineering}

One ubiquitous feature of our analysis that is often absent in smaller ecosystems is the large scale depletion of resources. As shown in Fig. \ref{CT-numeric1}, species can significantly change the resource profile and deplete a large fraction of
resources initially present in the environment. This environmental engineering  can change which species survive and thrive in an environment. One way to measure the effect of environmental engineering and the reshaping of the resource profile is to measure the correlation between the naive fitness of an organism, $f_i= \sum_{\alpha} c_{i \alpha} \max \{K_\alpha, 0\} -m_i$, and its steady-state abundance in the ecosystem $\overline{N}_i$. The fitness $f_i$ measures the growth rate of organism $i$ if it is introduced into an environment in the absence of other species. For this reason, we expect  $f_i$ to be highly predictive of $\overline{N}_i$ when resource abundance profiles are not significantly perturbed by consumption. On the other hand, in the presence of significant environmental engineering, we expect the correlation between $f_i$ and $\overline{N}_i$ to decrease significantly.

Fig. \ref{fig:Nversusf} shows  $f_i$ versus $\bar{N}_i$ for numerical simulations where the $c_{i \alpha}$ drawn from a binomial distribution with $p=0.1$ and $\sigma_m/m=0.1$, as well as the case where parameters are Gaussian random variables with mean an variance matching the binomial setting. From the figure, it is clear there is a significant correlation between $f_i$ and $\bar{N}_i$. Organisms with higher fitness disproportionately survive in the ecosystem. However, a significant number of organisms that have a high naive fitness $f_i$ can still go extinct in the ecosystem (black points). The difference between plots (A,B) and (C,D) is that in the former $K_\alpha$ and $m_i$ are kept positive by ensuring they are drawn from a gamma distribution and the consumer preferences $c_{i \alpha}$ are always positive since they are binary ($1$ with probability $p$ or $0$ otherwise). In (C,D), each of these parameters is drawn from a Gaussian distribution, but with the same mean and variance as in (A,B). This allows $K_\alpha$,
$m_i$, and $c_{i \alpha}$ to be negative. A negative $c_{i \alpha}$  means that species $i$ produces resource $\alpha$ at fitness expense to itself (i.e. public good production). As expected, this results in much more environmental engineering than the case where the $c_{i \alpha}$ are strictly positive. In C, red points indicate species with negative fitness that can stably exist in the community by utilizing public goods  ($f_i <0$, $\bar{N}_i>0$). In D, red points correspond to resources with a negative capacity which end up in the environment due to public good production by high abundance species. Conversely, species that cannot survive in the environment in the absence of other species can fixate due to environmental engineering (red points). Importantly, this emergent environmental engineering is a collective property of the whole ecosystem and results from a complex interplay between organisms and environment. These simulations demonstrate how environmental engineering can dramatically modify community structure.

Additionally, Fig.    \ref{fig:Nversusf} shows predictions from our CT for the correlation between $f_i$ and $\overline{N}_i$. Within our replica-symmetric ansatz, these correlations are described by normal distributions whose variances and covariances can be calculated using our self-consistent equations. The contour lines represent half a standard deviation spread of our normal distribution. Our theory qualitatively captures the shape of the correlation between $f_i$ and $\overline{N}_i$. We give explicit expression for these correlations as well as the mutual information between species abundances and naive fitness in Appendix \ref{sec:f_N_thy}.

\section{Discussion}

Niche-based theories have played a fundamental role in shaping our understanding of community assembly and community ecology. In this work, we use ideas and methods from statistical physics to analyze a canonical model in community ecology, MacArthur's Consumer Resource Model (MCRM). Unlike previous works, our statistical physics inspired approach allows us to analyze large ecosystems with many species and resources. Our results suggest that organisms can significantly perturb their environments.  The abundance of resources can be significantly altered and resource can even be completely depleted. We find that such niche-construction and environmental engineering is a generic feature of MCRM. This suggests that in complex ecosystems, organisms actively construct their environment. To quote Levins and Lewontin, ``they are not the passive objects of external forces, but creators and modulators of these forces" \cite{levins1985dialectical}. The effects of environmental engineering are even more dramatic when consumers can produce public goods at a fitness cost to themselves. In this case, species and resources that could not survive in isolation can fixate in the ecosystem.

To carry out our analysis, we developed a sophisticated theory based on the cavity method. One of the most striking things about our analysis is that many physical quantities that appear in the ``cavity equations'' have natural ecological interpretations in terms of invasion growth rates and effective carrying capacities.  The underlying reason for this is that the cavity methods is based on asking how ecosystems are perturbed when a new species and a new resource are introduced into the ecosystem. Conceptually, this is very similar in spirit to many classical arguments in community ecology pioneered by Levins and MacArthur that ask whether a new species can invade \cite{macarthur1967limiting, tilman1982resource}. This naturally allows us to generalize many of the results from niche-based theories to large, diverse ecosystems.  However, the price we pay for using our cavity approach is that we are limited to making statistical predictions.

An important question for future investigation is to ask how our results change if we make the model more realistic. In the MCRM, all species are assumed to have a linear, Type I functional response. It will be interesting to generalize our model to non-linear functional responses. We have also neglected the effects of environmental and demographic stochasticity. Stochasticity can induce phases transitions in ecosystems from a niche-like phase where competitive effects dominate community assembly to an ecologically neutral-like phase where stochasticity is the primary determinant of community structure \cite{fisher2014transition,dickens2016analytically}. It will be interesting to see if the techniques developed here can be generalized to this more complicated setting. Finally, we have assumed that our population can be modeled as a well-mixed community. However, spatial effects can qualitatively change the behavior of cellular populations \cite{korolev2010genetic,loreau1998biodiversity} and are likely to play an important role in community assembly.

\section{Acknowledgements}
We would like to thank Josh Goldford, Kirill Korolev, Seppe Kuehn, Alvaro Sanchez, Daniel Segr\`{e}, Cui Wenping for many useful discussions. PM was supported by  NIH NIGMS grant 1R35GM119461, a  Simons Investigator award in the Mathematical Modeling of Living Systems (MMLS), and a Scialog grant from the Simons Foundation and Research Corporation. MA was supported by the Swartz Program in Theoretical Neuroscience at Harvard.

\bibliography{refs_ecology}

\appendix

\section{Basic Setup}
We briefly summarize MacArthur's classical consumer resource model (MCRM). Species $i =1 \ldots S$ grows at a rate proportional to its utilization of resources, $R_\alpha$, $\alpha=1\ldots M$, in the environment.
This is described by the equation:
\be
{1 \over N_i} {d N_i \over dt} = \sum_{\alpha}  c_{i \alpha} R_{\alpha} -m_i +h_i,
\label{CRM_N}
\ee
 where $w_\alpha$ is the value of one unit of resource to a species (e.g. ATPs that can be extracted); $c_{i \alpha}$ is the rate at which species $i$ consumes resource $\alpha$ and converts that into a ``growth rate'', $m_i$; $m_i$ is the minimum amount of resources that must be consumed in order to have a positive growth rate. We have also added a small perturbation $h_i$ to the system that will do a linear expansion in. The original MCRM corresponds to the choice $h_i=0$.
 We define the growth rate to be
 \be
 g_i(\mathbf{R})= \sum_\alpha c_{i \alpha} w_\alpha  R_\alpha -m_i.
 \ee

In consumer resource model, resources satisfy their own dynamical equations:
\be
{d R_{\alpha} \over dt}= R_\alpha (K_\alpha +b_\alpha -R_\alpha)- \sum_j c_{j \alpha} N_j R_\alpha,
\label{CRM_R}
\ee
where the first term (with $b_\alpha =0$) describes the resource dynamics in the absence of any species and the second term models the consumption
of resource by species in the environment, and $b_\alpha$ is small perturbation. The original MCRM corresponds to the choose $b_\alpha=0$. Furthermore,
define the effective carrying capacity
\be
K_{\alpha}^{eff}=K_\alpha -\sum_j c_{j \alpha} N_j.
\ee

We will consider the case when the consumer preferences $c_{i \alpha}$ are random variables that can be characterized by their means and variances. In particular,
\be
\< c_{i \alpha}\>= {\mu_c \over S},
\ee
and
\be
\<c_{i \alpha} c_{j \beta}\>= {\sigma_c^2 \over S} \delta_{ij} \delta_{\alpha \beta} + {\mu_c^2 \over S^2} \approx {\sigma_c^2 \over S}.
\ee
To perform the cavity equations, it is useful to define several other quantities. Let us a define the fluctuating part of the consumer preferences $d_{i \alpha}$ as
\be
c_{i \alpha} \equiv {\mu_c \over S} + \sigma_c d_{i \alpha}.
\ee
Then, we have that
\be
\< d_{i \alpha}\>=0,
\ee
and
\be
\<d_{i \alpha} d_{j \beta} \>={\delta_{ij} \delta_{\alpha \beta} \over S}.
\ee

We will also assume that the carrying capacities are drawn from a Gaussian distribution with
\be
\< K_\alpha \>=K,
\ee
and
\be
\<\delta K_\alpha \delta K_\beta\>=\< (K_\alpha -K)(K_\beta-K)\>= \delta_{\alpha \beta} \sigma_K^2.
\ee
Finally, we assume that the minimum survival coefficients are also drawn from Gaussian distribution with
\be
\<m_i\>=m,
\ee
and
\be
\< \delta m_i \delta m_j\>= \<(m_i-m)(m_j-m)\>=\delta_{ij}\sigma_m^2.
\ee

For future reference, it will also be helpful to define the ratio
\be
\gamma={M \over S},
\ee
the average resource abundance,
\be
\<R\>= {1 \over M} \sum_{\alpha} R_\alpha,
\ee
and the average species abundance
\be
\<N\> = {1 \over S} \sum_j N_j.
\ee
With these definitions, notice that we can rewrite (\ref{CRM_N}) as
\be
{1 \over N_i} {d N_i \over dt} = \mu \gamma \<R\> -m +\sigma \sum_{\alpha} d_{i \alpha} R_\alpha - \delta m_i,
\label{CRM_N2}
\ee

and rewrite (\ref{CRM_R}) as
\be
{1 \over R_\alpha}{d R_{\alpha} \over dt}= [K -\mu_c \<N\>] -R_\alpha -\sigma_c \sum_j d_{j \alpha} N_j + \delta K_\alpha.
\ee
We can define the mean growth rate of the population
\be
g= \mu_c \gamma \<R\> -m
\ee
and the mean effective capacity of resources in the ecosystem to be
\be
K^{eff}=K -\mu_c \<N\>
\ee
 as in the main text.	In terms of these quantities, we can rewrite these equations as
\bea
{1 \over N_i} {d N_i \over dt} &=& g+\sigma \sum_{\alpha} d_{i \alpha} R_\alpha - \delta m_i, \nonumber \\
{1 \over R_\alpha}{d R_{\alpha} \over dt} &=& K^{eff}-R_\alpha -\sigma_c \sum_j d_{j \alpha} N_j + \delta K_\alpha.
\eea

The terms on the right hand sides of the equation above have a natural interpretation as the ``fluctuating parts'' of the growth rate and effective carrying capacity.
In particular, we have rewritten the growth rate for species $i$ as the sum of the mean growth rate $g$ and a fluctuating component  $\delta g_i$ defined
as
\be
\delta g_i =\sigma \sum_{\alpha} d_{i \alpha} R_\alpha - \delta m_i.
\ee
We have split the effective carrying capacity of resource $\alpha$ is divided into a mean $K^{eff}$ and fluctuating component $\delta K^{eff}_\alpha$ defined
as
\be
\delta K_\alpha ^{eff} = - \sum_j d_{j \alpha} N_j + \delta K_\alpha.
\ee

\section{Deriving the Species and Resource Distributions}

To derive the cavity equations, we will relate a system with $S$ species and $M$ resources to a new system where we add an additional
resource $R_0$ and and additional species $N_0$. Thus, the cavity equations relate a  ecosystem with $S+1$ and $M+1$ resources to a
ecosystem with $S$ and $M$ resources.

Then we can write equations for this new ecosystem (to leading order in $S$):
\be
{1 \over N_i} {d N_i \over dt} =g +\sigma_c \sum_{\alpha} d_{i \alpha} R_\alpha - \delta m_i + \sigma_c d_{i0} R_0,
\label{CRM_N0}
\ee
and
\be
{1 \over R_\alpha}{d R_{\alpha} \over dt}= K^{eff} -R_\alpha -\sigma_c \sum_j d_{j \alpha} N_j + \delta K_\alpha - \sigma_c d_{0 \alpha}N_0.
\label{CRM_R0}
\ee
We can also write down the corresponding equations for the new resource and species:
\be
{1 \over N_0} {d N_0 \over dt}= g +\sigma_c \sum_{\alpha} d_{0 \alpha} R_\alpha - \delta m_0 + \sigma_c d_{0 0} R_0,
\label{CEN}
\ee
and
\be
{1 \over R_0}{d R_0 \over dt}=K^{eff} -R_0 -\sigma_c \sum_j d_{j 0} N_j + \delta K_0 - \sigma_c d_{0 0}N_0.
\label{CER}
\ee

We now focus on steady-state. Let us denote the steady-state value of a quantity $X$ by $\bar{X}$. Then, we can define some susceptibilities
that are extremely useful for what follow:
\bea
\chi_{i\beta}^{(N)} &=& { \partial \bar{N}_i \over \partial K_\beta} \nonumber \\
\chi_{\alpha \beta}^{(R)} &=&  { \partial \bar{R}_\alpha \over \partial K_\beta}
\eea
and
\bea
\nu_{i j}^{(N)} &=& { \partial \bar{N}_i \over \partial m_j} \nonumber \\
\nu_{\alpha j}^{(R)} &=&  { \partial \bar{R}_\alpha \over \partial m_j}.
\eea

Now we are in a position to perform the cavity calculation. Let us denote the steady-state value of a quantity $X$ in the absence of the new resource and
species as $\bar{X}_{ /0}$. Then, since the addition of a resource and species represents a small perturbation (order $1/S$), we can write:
\be
\bar{N}_i= \bar{N}_{i /0} -\sigma_c  \sum_{\beta} \chi_{i\beta}^{(N)}d_{0 \beta} \bar{N}_0 -\sigma_c \sum_{j} \nu_{ij}^{(N)}d_{j0}\bar{R}_0,
\label{Nlinearresponse}
\ee
and
\be
\bar{R}_\alpha = \bar{R}_{\alpha /0} - \sigma_c \sum_{\beta} \chi_{\alpha \beta}^{(R)} d_{0 \beta} \bar{N}_0 -\sigma_c \sum_{j} \nu_{\alpha j}^{(R)}d_{j0}\bar{R}_0.
\label{Rlinearresponse}
\ee
We can now plug in these expressions into the steady-state equations for $N_0$ and $R_0$. This gives:
\begin{widetext}
\be
0=\bar{N}_0 \left[g + \sigma d_{0 0} R_0 + \delta m_0 +\sigma_c  \sum_{\alpha} d_{0 \alpha} R_{\alpha /0} -\sigma_c^2 \sum_{\alpha \beta} \chi_{\alpha \beta}^{(R)}
d_{0 \alpha} d_{0 \beta} N_0 -\sigma_c^2 \sum_{\alpha , j} \nu_{\alpha j}^{(R)} d_{0 \alpha} d_{j 0} R_0 \right].
\ee
If we now take leading order contributions to $S$ in this expression, and take expectation value over expressions this reduces to
\be
0=\bar{N}_0 \left[g+ \sigma d_{00} R_0   -\gamma \sigma^2 {1 \over M} \sum_{\alpha} \chi_{\alpha \alpha}^{(R)} N_0 + \delta m_0 +\sigma  \sum_{\alpha} d_{0 \alpha} R_{\alpha /0}\right].
\ee
\end{widetext}
Notice that, to leading order in $S$, we can model the term $ \delta m_0 +\sigma  \sum_{\alpha} d_{0 \alpha} R_{\alpha /0}$, which is just the invasion growth rate minus the mean
growth rate $g_0^{inv}-g$, as a Gaussian random field with mean $0$ and variance:
\be
\sigma_g^2=  \sigma_c^2 \gamma  {1 \over M} \sum_{\alpha} \bar{R}^2_{\alpha / 0} + \sigma_m^2 = \sigma_c^2 \gamma q_R + \sigma_m^2,
\label{defsigmaN}
\ee
where
\be
 q_R ={1 \over M}  \sum_{\alpha} R_{\alpha /0}^2 .
\ee

If we let $z_N$ be random field with mean zero and unit variance, and define the average suspectibility
\be
\chi= {1 \over M} \sum_{\alpha} \chi_{\alpha \alpha}^{(R)}
\ee
then we can write the equation for $N_0$ as
\be
0=\bar{N}_0 \left[\mu \gamma \<R\> -m  -\gamma \sigma^2 \chi \bar{N}_0 + \sigma_g z_N \right].
\ee

We can also derive a similar equation for $R_0$. This is given by
\begin{widetext}
\be
0= R_0 \left[K^{eff} -R_0 -\sigma_c \sum_j d_{j 0} N_{j /0} -\sigma_c^2 \sum_{j  \beta} \chi_{i\beta}^{(N)} d_{j 0} d_{0 \beta} \bar{N}_0
-\sigma_c^2  \sum_{j, k} \nu_{j k}^{(N)} d_{j 0} d_{k 0} \bar{R}_0+ \delta K_0 - \sigma_c d_{00}N_0 \right].
\ee
\end{widetext}
Using the same logic as above, to leading order we have
\be
0= R_0 \left[K^{eff} -R_0 +\sigma_c^2 \nu \bar{R}_0 +\sigma_{K^{eff}} z_R \right],
\ee
where
\be
\nu = {1 \over S} \sum_j \nu_{j j}^{(N)},
\ee
and
\be
\sigma_{K^{eff}}^2 = \sigma_K^2+ \sigma_c^2 q_N,
\label{defsigmaR}
\ee
with
\be
 q_N ={1 \over S}  \sum_{j} N_{j /0}^2 .
\ee

We can solve these equations and get
\bea
N_0  &=& { \max \left[ 0,  {g + \sigma_g z_N }\right] \over  \gamma \sigma^2 \chi} \\
R_0 &=& { \max \left[0, K^{eff} + \sigma_{K^{eff}} z_R \right] \over 1-\sigma^2 \nu}.
\label{EqNR}
\eea
Thus, the distributions for $N$ and $R$ are given by truncated Gaussians.

\section{Self Consistency Equations}
\label{sec:self-con}
Let us now write some self-consistency equations in the replica symmetric phase. Let us define the number of non-zero species and resources as $S^*$ and $M^*$ respectively. Furthermore, define
\bea
\phi_S &=& {S^* \over S}, \\
\phi_M &=& {M^* \over M}.
\eea
Our goal is, given some parameters $\{ K, \sigma_K, m, \sigma_m, \mu, \sigma, S,M\}$, to find the values for $\{ \phi_S, \phi_N, \<N\>, \<R\>, q_R, q_N, \chi, \nu\}$. Since there are eight unknowns we will need eight equations. It will also be useful to define:
\bea
\Delta_g &=&  {\mu \gamma \<R\>-m \over \sigma_g}, \\
\Delta_{K^{eff}} &=& {K-\mu\<N\> \over \sigma_{K^{eff}}},
\label{DefDelta}
\eea
and the function
\be
w_j(\Delta)= \int_{-\Delta }^\infty {dz \over \sqrt{2 \pi}} e^{-z^2 \over 2} (z+\Delta)^j.
\ee
First, let us write self-consistency equations for the susceptibilities. Taking derivatives with respect to $m$ and $K$ of (\ref{EqNR}) and noting that the fraction  of non-zero species and non-zero resources is $\phi_N$ and $\phi_R$ respectively gives
\bea
\nu = -{\phi_N \over \gamma \sigma_c^2 \chi}, \\
\chi= {\phi_R  \over 1- \sigma_c^2 \nu}.
\label{SupSCE}
\eea
Notice now that if we define  $y= \max[0,{a \over b} + {c \over b} z]$ with $z$ a gaussian random variable we have that:
\be
\<y^j\>= \left({b \over c}\right)^j \int_{-{b \over a}}^ \infty dz {1 \over \sqrt{2 \pi}} e^{-z^2 \over 2}(z+{b \over a})^j.
\ee

We can now use the fact that (\ref{EqNR}) implies that the species distribution and resource distribution is given by a truncated Gaussian to write self consistency equations for the fraction of nonzero resources and species as well as the the moments of their abundances:
\bea
\phi_N = w_0(\Delta_g), \label{eq:phi_N}\\
\phi_R = w_0(\Delta_{K^{eff}}),\label{eq:phi_R}\\
\<N\>=  \left({ \sigma_g \over \gamma \sigma_c^2 \chi}\right)w_1(\Delta_g),\\
\<R\>=\left({ \sigma_{K^{eff}} \over 1-\sigma_c^2 \nu}\right)w_1(\Delta_{K^{eff}}), \label{eq:avg_R}\\
q_N=\<N^2\>=\left({ \sigma_g \over \gamma \sigma_c^2 \chi}\right)^2 w_2 (\Delta_g),\\
q_R=\<R^2\>=\left({ \sigma_{K^{eff}} \over 1-\sigma_c^2 \nu}\right)^2 w_2(\Delta_{K^{eff}}).
\label{ExSCE}
\eea
Together (\ref{DefDelta}), (\ref{SupSCE}), and (\ref{ExSCE}) define the 10 self-consistency equations we need, along with the definitions (\ref{defsigmaN}) and (\ref{defsigmaR}).

We solve these mean field equations numerically using the sum of squared differences between the left and right sides of equations (\ref{eq:phi_N}-\ref{ExSCE}) as an energy function which we minimize using the basinhopping optimization algorithm from the scipy.optimize. The algorithm uses random perturbations, local minimization, and an accept or reject criterion to attempt to minimize function which may be non-convex. The parameters we used were a temperature of $1$, a step size of $0.5$, and $5$ iterations or initializations. Note that the equations can also be solved iteratively, but we found these solutions were stable for a smaller set of parameter values using this approach.

\section{Zero net-growth nullclines and generalized impact vectors}

We can also easily relate our mean-field quantities to ZGNIs. Recall, that ZGNI's  delineate range of resource conditions in which a species maintains a positive growth rate \cite{chase_ecological_2003, letten2017linking}. Each species  $i$ defines a ZGNI in the resource space $\mathbf{R}_i^{ZGNI}$  defined by the equation $g_i (\mathbf{R}_i^{ZGNI})=0$. Geometrically, we can view $\mathbf{R}_i^{ZGNI}$ as a hyperplane in resource space whose dot product with the consumption coefficients $c_{i \alpha}$ of species $i$  equals $m_i$ (see Eq. 1). If $m_i \ll 1$, the ZGNI is well-approximated by the plane perpendicular to $c_{i \alpha}$.  We can calculate some statistical properties of these ZGNI. Notice that the mean value of each component is just
\be
\<R_{i \alpha}^{ZGNI}\>= {m \over \gamma \mu_c},
\ee
and the expected value of the square is just
\be
\<(R_{i \alpha}^{ZGNI})^2 \>= {\sigma_m^2 + m^2 \over \gamma \sigma_c^2}.
\ee
In Modern Niche Theory, another important quantity is the impact vector of a species $i$. The impact vector describes how resources are depleted by the addition of another individual. Here, we introduce the idea of generalized impact vectors that measure how the steady state concentration of a resource $\alpha$ changes due to the introduction of a species $j$  Alternatively, we can consider a system without resource $\beta$ and then ask how it's addition changes the species abundance of a species $i$ . These define the generalized impact vectors (GIVs).

These are of course the leading order contribution (in $S$)  to the cavity equations under the replica symmetric assumption, namely (\ref{Rlinearresponse}) and (\ref{Nlinearresponse}).
Thus, the components of the two ``generalized' impact vectors are given by:
\be
\bar{R}_\alpha - \bar{R}_{\alpha /j}=  - \sigma_c \sum_{\beta} \chi_{\alpha \beta}^{(R)} d_{j\beta} \bar{N}_j,
\ee
and
\be
\bar{N_i}- \bar{N}_{i /\beta} = -\sigma_c \sum_{j} \nu_{ij}^{(N)}d_{j\beta}\bar{R}_\beta.
\ee

\section{Comparison of individual fitness to true growth rate and steady state abundance}
\label{sec:f_N_thy}
\noindent We want to quantify how much the naive fitness (the growth rate of an organism without other species present) is correlated to the invasion growth rate $g_i^{inv}$, which in turn is closely related to the steady state abundance of each of the species in the community. From the definition of the invasion growth rate:
\begin{equation}
g_i^{inv} = \sum_\alpha{c_{i\alpha}R_{\alpha / i}} - m_i,
\end{equation}
and naive fitness:
\begin{equation}
f_i = \sum_\alpha{c_{i\alpha}K_\alpha} - m_i,
\end{equation}
we can compute the level of correlation between the two using the CT. To begin, we compute the means of each of these distributions:
\begin{equation}
\<f_i\> = \sum_\alpha{\<c_{i \alpha}\> \<K_\alpha\>} - \<m_i\> = \gamma \mu_c K - m,
\end{equation}
and
\begin{equation}
\<g^{inv}_i\>  =  \sum_\alpha{ \<c_{i \alpha}\> \<R_{\alpha/i}\>} - \<m_i\>  = \gamma \mu_c \< R \> - m.
\end{equation}
Note that the consumer preferences $c_{i \alpha}$ of an individual species are independent of the resources steady state levels when that species $i$ is not included in the community as in the previous equation. We can also compute the correlation between fluctuations from the mean naive fitness and the mean invasion growth rate: i.e. if a species has a higher or lower individual fitness, how should we expect this to impact its growth rate in the community?
To understand this correlation, we define $\delta f_i = f_i - \< f_i\>$ and $\delta g^{inv}_i = g^{inv}_i - \< g^{inv}_i\>$ and compute:
\begin{eqnarray*}
  \<\delta f_i \delta g^{inv}_i\> &=& \< \left(\sum_\alpha{\left[\sigma_c d_{i \alpha} K_\alpha + \frac{\mu_c}{S} (K_\alpha -\<K_\alpha\>)\right]} + \delta m_i \right) \\
  & \cdot & \left(\sum_\alpha{\left[\sigma_c d_{i \alpha} R_{\alpha/i} + \frac{\mu_c}{S} (R_{\alpha/i} -\<R_{\alpha/i}\>)\right]} + \delta m_i \right)  \>.
\end{eqnarray*}
In the large $S$ limit, the important terms remaining in the average above are:
\begin{equation}
  \<\delta f_i \delta g^{inv}_i\> = \gamma \frac{\sigma_c^2}{M} \sum_\alpha{R_{\alpha/i} K_\alpha} + \sigma_m^2,
\end{equation}
thus in the asymptotic limit we can write the correlation between the two forms of fitness as:
\begin{equation}
\<\delta f_i \delta g^{inv}_i\> = \gamma \sigma_c^2 \< R_{\alpha/i} K_\alpha \> + \sigma_m^2.
\end{equation}
To compute the correlation between carrying capacity and resource level we modify \eqref{EqNR} from our capacity calculation, which yields such a relationship:
\begin{equation}
R_\alpha(K_\alpha) = { \max \left[0, K_\alpha-\mu_c\<N\> + \sigma_c \sqrt{q_N} z_N\right] \over 1-\sigma_c^2 \nu}.
\end{equation}
Using this relationship and letting $k$ be drawn from the same distribution as $K_\alpha$, where
 %along with the fact that the change in $\< N\>$ due to variations in $K_a$ will not be leading order in $\frac{1}{S}$ in what follows. We take the functional form
\begin{equation}
R(k) =  { \max \left[0, k-\mu_c\<N\> + \sigma_c \sqrt{q_N} z_N\right] \over 1-\sigma_c^2 \nu},
\end{equation}
we compute
\begin{equation}
\< k R(k) \>_{k, z_N}.
\end{equation}
The full form of this integral is thus:
\begin{equation}
\<\int{dk  k R(k) e^{-\frac{(k - K)^2}{2 \sigma_K^2} }  }\>_{z_N}.
\end{equation}
By rewriting this as
\begin{equation}
\<\int{dk  (k - K )R(k) e^{-\frac{(k - K)^2}{2 \sigma_K^2} }  } + \int{dk  K R(k) e^{-\frac{(k - K)^2}{2 \sigma_K^2} }  }\>_{z_N},
\end{equation}
it may be simplified via integration by parts on the first term, yielding:
\begin{eqnarray*}
  \< k R(k) \>_{k, z_N} &=& \sigma_K^2 \<R'(k)\> + K \< R(k)\> \\
   &=& \frac{1}{1 - \sigma^2 \nu } \left(\sigma_K^2 w_0(\Delta_{K^{eff}}) + K w_1(\Delta_{K^{eff}})\right),
\end{eqnarray*}
where $\Delta_{K^{eff}} = \frac{K - \<N\>}{\sigma_{K^{eff}}}$, and
\be
w_j(\Delta)= \int_{-\Delta }^\infty {dz \over \sqrt{2 \pi}} e^{-z^2 \over 2} (z+\Delta)^j.
\ee
Note, we can also compute Pearson's correlation coefficient for these two fitness metrics:
\begin{eqnarray}
  % Remove numbering (before each equation)
   c_\rho &=& \frac{\<\delta f \delta g^{inv}\>}{\sqrt{\<(\delta f)^2 \> \<(\delta g^{inv})^2 \>}} \\
   &=& \frac{\gamma \sigma_c^2 \< k R(k)\>_{k, z_N} + \sigma_m^2}{\sqrt{(\sigma_m^2+\gamma \sigma_c^2 (\sigma_K^2 + K^2))(\gamma \sigma_c^2 \<R^2\> + \sigma_m^2)}} \\
   &=& \frac{\frac{\gamma \sigma_c^2}{1 - \sigma_c^2 \nu } \left(\sigma_K^2 w_0(\Delta_{K^{eff}}) + K w_1(\Delta_{K^{eff}})\right) + \sigma_m^2}{\sqrt{(\sigma_m^2+\gamma \sigma_c^2 (\sigma_K^2 + K^2))(\gamma \sigma_c^2 \<R^2\> + \sigma_m^2)}}\nonumber.\label{eq:crho}
\end{eqnarray}
Using \eqref{eq:phi_R} and $\eqref{eq:avg_R}$ we can write the preceding expression as:
\begin{equation}
\frac{\frac{\gamma \sigma_c^2}{1 - \sigma_c^2 \nu } \left(\sigma_K^2 \phi_R + K  \left(\frac{1 - \sigma_c^2 \nu}{\sigma_{K^{eff}}}\right)\<R\>\right) + \sigma_m^2}{\sqrt{(\sigma_m^2+\gamma \sigma_c^2 (\sigma_K^2 + K^2))(\gamma \sigma_c^2 \<R^2\> + \sigma_m^2)}}.
\end{equation}

\subsection{Abundance vs naive fitness}

Note that given the relationship that $N$ is a scaled version of $g^{inv}$ where all negative values are truncated to zero \eqref{EqNR}, it follows that we can compute the correlation between $\tilde{N}_i$ and $f_i$, where $\tilde{N}_i = \frac{g^{inv}_i}{\gamma \sigma_c^2 \chi}$. Where we let $N_i = \tilde{N}_i$ if $\tilde{N}_i>0$ and $N_i = 0$ otherwise. To better understand that the correlation between the abundance and fitness of a species, we compute the correlation between $\tilde{N}$ and $f$:
\begin{equation}
\small{
\<\delta \tilde{N} \delta f \>=  \frac{1}{\gamma \sigma_c^2 \chi}\left(\frac{\gamma \sigma_c^2}{1 - \sigma_c^2 \nu } \left(\sigma_K^2 \phi_R + K  \left(\frac{1 - \sigma_c^2 \nu}{\sigma_{K^{eff}}}\right)\<R\>\right) + \sigma_m^2 \right)},
\end{equation}
\begin{equation}
\<(\delta \tilde{N})^2\>= \frac{(\gamma \sigma_c^2 \<R^2\> + \sigma_m^2)}{(\gamma \sigma_c^2 \chi)^2},
\end{equation}
and
\begin{equation}
\< (\delta f )^2\> = \sigma_m^2+\gamma \sigma_c^2 (\sigma_K^2 + K^2).
\end{equation}
Using these covariances, along with the means:
\begin{equation}
\<f\> = \gamma \mu_c K - m,
\end{equation}
and
\begin{equation}
\<\tilde{N}\>  = \frac{\gamma \mu_c \< R \> - m}{\gamma \sigma_c^2 \chi},
\end{equation}
we are able to generate theoretical predictions for the distribution of $f,N$. See Fig \ref{fig:Nversusf} where the theoretical plot of $f$, $\tilde{N}$ is compared with values of fitness $f$ and abundance $N$ for all species in a network over many realizations.

\subsection{Resource capacity versus resource abundance}
In the same Fig \ref{fig:Nversusf}, we additionally plot a theoretical prediction overlayed with numerics of how the resource capacities $K_\alpha$ are related to the resource abundances $R_\alpha$.

If we define $\tilde{R}_\alpha = \frac{K^{eff}_\alpha}{1 - \sigma_c^2 \nu}$, which is equal to the prediction of $R_\alpha$ when the resources abundance is positive, and use the fact
\begin{equation}
\delta K^{eff}_\alpha = - \sum_j d_{j \alpha} N_{j} + \delta K_\alpha,
\end{equation}
then we may combine these two relations to yield:
\begin{equation}
\delta \tilde{R}_\alpha = \frac{1}{1 - \sigma_c^2 \nu}\left( -\sum_j d_{j \alpha} N_{j/\alpha} + \delta K_\alpha\right).
\end{equation}
We can thus compute the correlations:
\begin{equation}
\frac{1}{M}\sum_\alpha{\delta K_\alpha \delta K_\alpha} \ra \sigma_K^2,
\end{equation}

\begin{equation}
\frac{1}{M}\sum_\alpha{\delta K_\alpha \delta \tilde{R}_\alpha} \ra \frac{\sigma_K^2}{1 - \sigma_c^2 \nu},
\end{equation}

\begin{equation}
\frac{1}{M}\sum_\alpha{\delta \tilde{R}_\alpha \delta \tilde{R}_\alpha} \ra \frac{\sigma_c^2 q_N + \sigma_K^2}{(1 - \sigma_c^2\nu)^2}.
\end{equation}
Also the means are easy to compute:

\begin{equation}
\<K_\alpha\> \ra K,
\end{equation}

\begin{equation}
\<\tilde{R}_\alpha\> \ra \frac{K - \mu \<N\>}{1 - \sigma_c^2\nu}.
\end{equation}
This allows us to make theoretical predictions for how the resource abundances are correlated to resource capacities.

%\begin{equation}
%  c_\rho = \frac{\<\delta f \delta g^{inv}\>}{\sqrt{\<(\delta f)^2 \> \<(\delta g^{inv})^2 \>}}= \frac{\gamma \< R K\>}{\sqrt{(\sigma_m^2+\gamma \sigma_c^2 (\sigma_K^2 + K^2))(\gamma \<R^2\> + \sigma_m^2)}}.
%\end{equation}

\begin{figure}
\includegraphics[width=1.0\columnwidth]{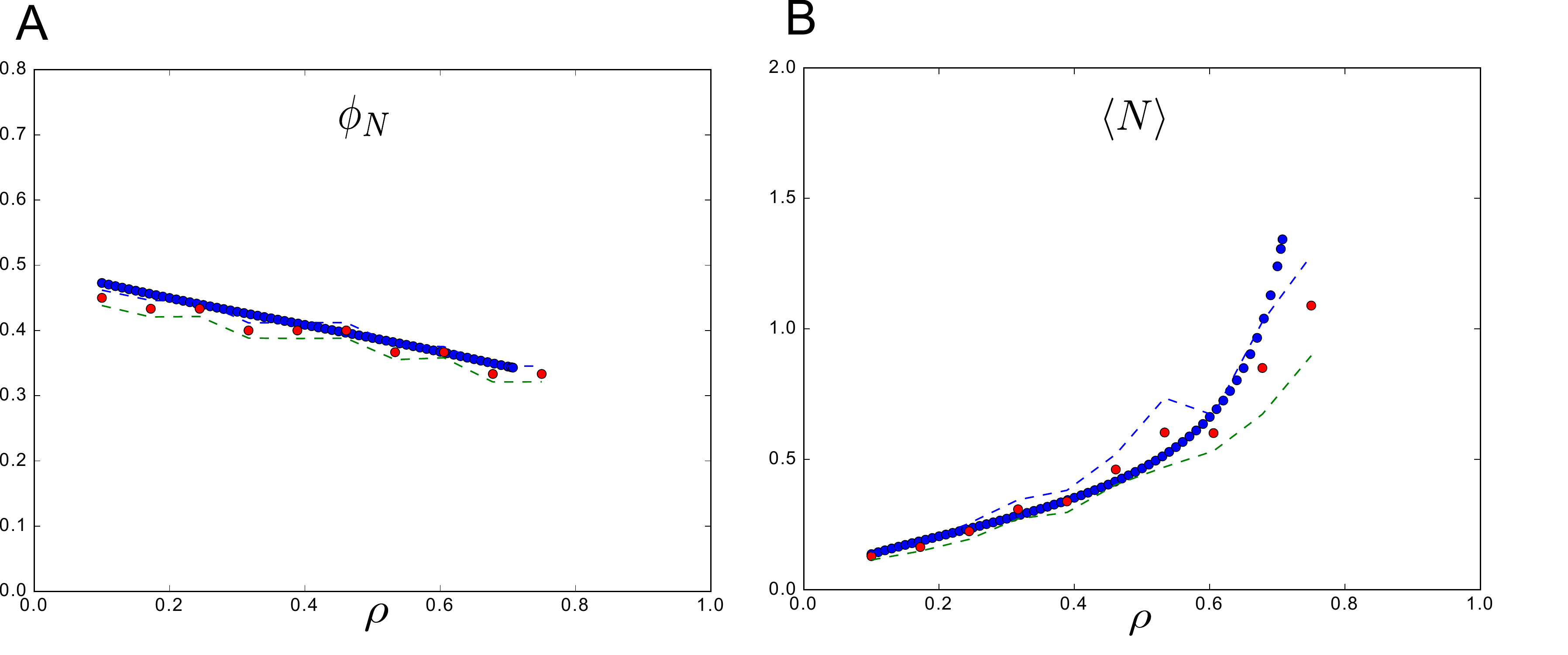}
\caption{{\bf Comparing CRM theory vs simulation - another setting}. All parameters defining the species were drawn from a Gaussian distribution as in main text but with $\sigma_m=1$ (a larger maintenance cost variance). Interestingly, we see a different behavior than in Fig \ref{CT-numeric1} in that the average abundance $\<N\>$ of all species is increasing with increasing niche overlap. Also, the number of surviving species is reduced by high niche overlap as it is in Fig \ref{CT-numeric1}, which makes sense since when species are very similar they will compete more directly for the same resources. The parameters used are: $\sigma_m =1, m = 1, \sigma_K = 1, K = 1$, and simulations were run assuming an equal number of species and resources $S = M = 30$ ($\gamma = 1$). Additionally, $\mu_c = 1$ was fixed and $\sigma_c$ varied (note that the niche overlap $\rho$ is determined by these two variables). The theoretical predictions from the cavity solution (blue), match well with the numerical solutions (red) for the CRM model averaged over 100 trials with and plotted with $\pm 2$ standard deviation error bars (dashed lines).
}
\label{fig:app_gauss_crm}
\end{figure}

\subsection{Mutual information between individual fitness and true growth rate}

The mutual information between two Gaussian variables $x$ and $z$ is simply (note the means of these random variables do not contribute so we will assume them are zero mean random variables):
\begin{equation}
I(x,z) = \frac{1}{2}\ln \left(\frac{\sigma_x^2 \sigma_z^2}{\sigma_x^2 \sigma_z^2 - \sigma_{xz}^2} \right).
\end{equation}
Thus,
\begin{equation}
I(f, g^{inv}) = -\frac{1}{2}\ln \left(1 - c_\rho^2\right).
\end{equation}
This gives us a theoretical prediction using the predicted form for the correlation coefficient $\eqref{eq:crho}$.
%\begin{equation}
% c_\rho = \frac{\<\delta f \delta g^{inv}\>}{\sqrt{\<(\delta f)^2 \> \<(\delta g^{inv})^2 \>}} = \frac{ \frac{\gamma}{1 - \sigma^2 \nu } \left(\sigma_K^2 w_0(\Delta_{K^{eff}}) + \mu_k w_1(\Delta_{K^{eff}})\right) + \sigma_m^2}{\sqrt{(\sigma_m^2+\gamma \sigma_c^2 (\sigma_K^2 + K^2))(\gamma \<R^2\> + \sigma_m^2)}}.
%\end{equation}

\section{Additional simulations and notes}

We discussed how the theoretical curves were generated in Appendix \ref{sec:self-con}. The numerical simulations were performed by solving the corresponding ODEs \eqref{Eq:MCRM} and integrating numerically until time $50,000$ with $1000$ steps. Although it is not always needed, we improved the accuracy by additionally including a small amount of migration noise which we lowered linearly to a negligible roundoff error over the course of the integration to help ensure that a species that was favored to survive would not go extinct.

We also ran simulations in other regimes, such as the one shown in Fig \ref{fig:app_gauss_crm} where we consider fixing $\mu_c = 1$ while varying $\sigma_c$ to study the setting when we are less interested in comparing specialists to generalists and more interested in the effect of niche overlap and how a high overlap in the generating distribution can reduce the number of surviving species.

\end{document}